\documentclass[english]{article}
\usepackage[T1]{fontenc}
\usepackage[latin9]{inputenc}
\usepackage{geometry}
\usepackage{array}
\usepackage{url}
\usepackage{amsmath}
\usepackage{stackrel}
\usepackage{graphicx}
\usepackage[authoryear]{natbib}

\makeatletter

\providecommand{\tabularnewline}{\\}

\usepackage{mathptmx}
\usepackage[symbol]{footmisc}

\makeatother

\usepackage{babel}
\begin{document}
\title{\textsc{GotFunding}: A grant recommendation system based on scientific
articles}
\author{Tong Zeng$^{1,2,}$ , Daniel E. Acuna$^{2,}$\thanks{Corresponding author: \protect\url{deacuna@syr.edu}}}
\date{$^{1}$School of Information Management, Nanjing University, Nanjing
210023, China\\
$^{2}$School of Information Studies, Syracuse University, Syracuse,
NY 13244, USA}
\maketitle
\begin{abstract}
Obtaining funding is an important part of becoming a successful scientist.
Junior faculty spend a great deal of time finding the right agencies
and programs that best match their research profile. But what are
the factors that influence the best publication--grant matching?
Some universities might employ pre-award personnel to understand these
factors, but not all institutions can afford to hire them. Historical
records of publications funded by grants can help us understand the
matching process and also help us develop recommendation systems to
automate it. In this work, we present \textsc{GotFunding} (Grant recOmmendaTion
based on past FUNDING), a recommendation system trained on National
Institutes of Health's (NIH) grant--publication records. Our system
achieves a high performance (NDCG@1 = 0.945) by casting the problem
as learning to rank. By analyzing the features that make predictions
effective, our results show that the ranking considers most important
1) the year difference between publication and grant grant, 2) the
amount of information provided in the publication, and 3) the relevance
of the publication to the grant. We discuss future improvements of
the system and an online tool for scientists to try.
\end{abstract}

\section{Introduction}

The ability of a scientists to fund themselves plays an important
role in a scientist's career, sometimes propelling their productivity
\citep{jacob2011impact}. Scientists, thus, spend an enormous amount
of time finding the right opportunities, writing proposals, and waiting
for funding decisions \citep{herbert2013time}. Past researchers have
estimated that the opportunity costs in searching and preparing a
grant might not be worth it \citep{gross2019contest}. Some solutions
to this problem include less stringent criteria for junior faculty
\citep{van2015early}, awarding grants with a lottery \citep{gross2019contest},
or peer-funding mechanisms \citep{bollen2014funding}. Here we explore
yet another alternative that instead uses machine learning to suggest
the best-matching grant based on her publications. We show that we
can cast the problem as a recommendation system trained on historical
grant--publication data. Our work attempts to improve funding success
which plays such a crucial role in today's careers.

Finding the right grant is important and there are several factors
involved in it. Scientists usually need to juggle multiple criteria
including funding agencies (e.g., NSF or NIH), career stages (e.g.,
junior-oriented or senior/leader-oriented), award amounts (e.g., small
NSF grant vs large DARPA grant), funding lengths (e.g., 1-year EAGER
NSF grant or 5-year CAREER NSF grant), and call relevance (e.g., a
particular program within NSF or institute in NIH) \citep{li2012having}.
Thousands of grant opportunities might be available at any given time,
offering hundreds of millions of dollars combined \citep{boroush2016us}.
These opportunities also have ramifications far beyond the receipt's
career \citep{lane2009assessing}. It is therefore hard to navigate
these funding opportunities but there should be ways in which to improve
the process.

Several  researchers have proposed numerous ways to improve the grant
review process. In the work of \citet{bollen2014funding}, the authors
proposed that funding agencies could distribute funding equally during
a first round, and, in subsequent rounds, scientists could send a
portion of this funding to other researchers that they think deserve
the funding. In a more recent work, \citet{gross2019contest} proposed
a mechanism where grants that pass a certain (low) decision threshold
go through a lottery mechanism. In simulation, the authors showed
that scientist itself benefits more because scientists spend more
time doing actual research than preparing grants. These methods, however,
are not considering that perhaps scientists are not applying to the
best-matching funding opportunities. Thus, the present study provides
a solution to improving the current state of affairs.

While submitting a grant is time consuming and has low probability
of success (e.g., see \citet{gross2019contest,bollen2014funding}),
these low probabilities might be related to a mismatch between the
grant submitted and the agency that receives it \citep{crow2020your}.
Another way of improving the granting process is rather than changing
the preparation and review process, we could improve the quality of
the matching between scientists and opportunities. Recommendation
systems are a natural way of improving how scientists find relevant
information such as publications (e.g., \citet{achakulvisut2016science}).
A similar process could be applied to grant recommendation systems.
Some systems exists (e.g., Elsevier's Mendeley Funding \citet{mendeleyfunding})
but they are closed source and difficult to evaluate. Thus, the granting
process can be improved by increasing the submission accuracy using
recommendation systems.

In this publication, we propose to use historical data of past publication--grant
relationships from NIH. We cast the problem as a learning-to-rank
recommendation system and show that it can achieve high performance
on validation (NDCG@1 = 0.945). We further explore the factors that
maximize the quality of the match, suggesting that successful scientists
match publications to temporarily relevant grants and achieve high
publication--grant match relevance. We describe potential improvements
in the future.

\section{Materials and Methods}

\subsection{Recommendation as Ranking}

\begin{figure}
\begin{centering}
\includegraphics[scale=0.4]{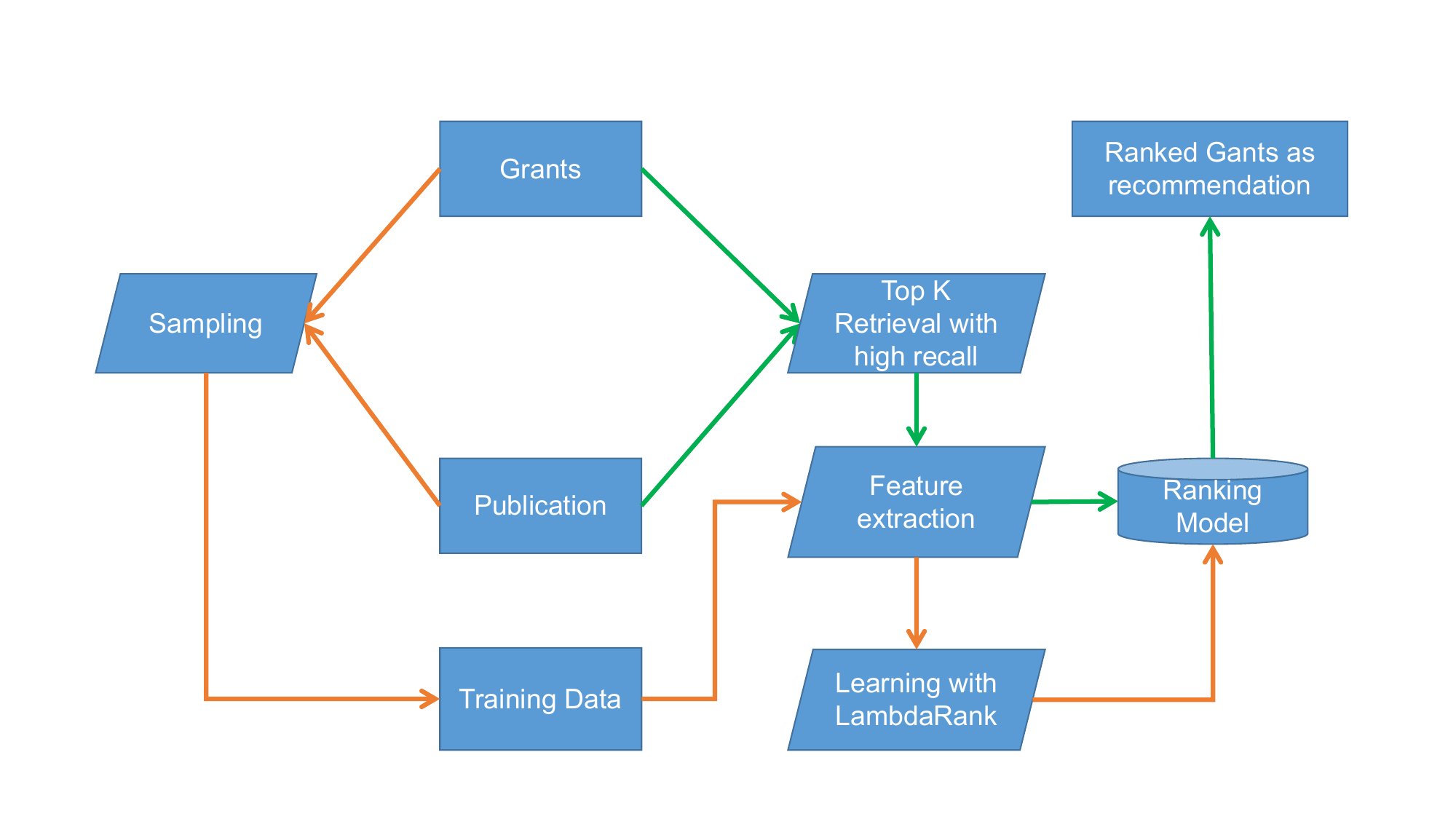}
\par\end{centering}
\caption{\protect\label{fig:framework}The framework of our grant recommendation
solution. The orange arrows denote the training pipeline and the green
arrows represent the prediction pipeline.}
\end{figure}

Suppose a user has associated a set of publications $P$ where each
publication contains year and some description, such the title and
abstract. These publications could be submitted by the user, or based
on the user browsing history, or come from the user's publication
history. Also, there are announcements/messages/notifications from
the funding agencies stating new grants and calling for proposals,
which we denote as funding opportunities, $G=\{g_{1},g_{2},...g_{k}\}$,
where each funding opportunities $g_{i}$ contains information such
as funding description, year, and agency. Our grant recommendation
solution could be defined as using the $P$ as input, producing a
subset of $G$ as output $R$ (\emph{retrieval stage}), ranking each
item in $R$ by the relevance value between the publication and opportunity,
and returning funding candidates ranked by relevance (\emph{ranking
stage}). The overall framework is shown in Figure \ref{fig:framework}.

Since there is already mature solution for retrieval stage, as an
exploratory research, we are focus on the learning an effective ranking
function. In the ranking stage, we need a function to assign a matching
score to each retrieved grant candidate. The ranking order based on
these scores indicates the relevance between the grants candidates
and the publication. Learning such a ranking function is an important
task in machine learning, called learning-to-rank. Depending on how
the loss function is optimized, learning to rank can be categorized
into pointwise, pairwise and listwise approaches \citep{l2r_listwise,li2011short,burges2010ranknet}.
For pointwise approach, the loss function takes only one document
into account and optimizes to predict the relevant score directly.
The pairwise ranking inputs a pair of documents into the loss function,
and minimizes the incorrect ranking of these two documents compared
to the ground truth. The listwise method looks at the candidate list
directly, and tries to find the optimal ordering. In practice, the
pairwise is more accurate than pointwise approach, and the list-wise
approach is much more complex compared to the point-wise and pair-wise.
In this paper, we will use pair-wise approach. Specifically, we will
use the LambdaRank algorithm \citep{l2r_pairwise} implemented by
lightGBM \citep{lightgbm}.

\subsection{Datasets}

\subsubsection{Federal RePORTER}

Federal RePORTER is an open and automated data infrastructure that
collects data on federally funded research projects and its outcomes
(e.g. publications and patents). The federal RePORTER includes approximately
1.15 million projects from 2000 to 2019, and involving 18 agencies.
Among all the agencies, the NIH accounts for 77.3\% of all the projects
and it has the biggest funding pool\footnote{see https://federalreporter.nih.gov/Home/FAQ\#faqs-panel7 for the
projects distribution over agencies}. In this publication, we focus only on NIH publication--grant relationships.
Each of the NIH projects contains a list of the publications acknowledging
the grant. Most of this publications are from PubMed, which we now
describe.

\subsubsection{PubMed}

PubMed is a search engine and publication repository developed and
maintained by the United States National Library of Medicine (NLM)
at the NIH and mainly focuses on the fields of biomedical and health
science. It provides access to over 30 million publications from MEDLINE
(an NLM journal citation database), life science journal and online
books. We use this publications in our recommendation system. We downloaded
the 2019 baseline and the subsequent daily updates on December 2019.

\subsubsection{Statistics of the datasets}

We perform some data filtering and cleaning, such as removing duplication,
removing projects and publications without links in Federal RePORTER
table. We removed these sub-grants. Further, we removed outliers such
as grants that yield more than 10 publications and publications which
are funded by more than 3 grants. In the end, we have 67,396 grants
and 235,419 publications.

\subsubsection{Training data for learning to rank}

The recommendation system learned from training data that starts with
a list of publications. We create an artificial ranking using the
following scheme. Rank 1 are grants that actually funded a publication.
Rank 2 is the nearest neighbor grant. Rank 3, 4, and 5 are the first,
second, and third distance quantile to the publication. The distance
measure used is cosine similarity tf-idf vector space. This initial
data is therefore an list of ordered lists, one for each publication,
containing five grants each. Using these lists, we then proceed to
extract features that can be used to learn the ranking.

\subsection{Learning Features Extraction}

We concatenate the publication title and abstract as the publication
description. We consider the grant descriptions as the following fields:
1) the funding agency information (e.g., full name and description),
2) the grant's title, 3) grant's abstract, and 4) the union of 1 through
3. For each grant-publication pair, we extract the statistical and
semantic features described in the next section.

\subsubsection{For Statistical features}

For each grant-publication pair, we extract 31 statistical features
(see Table \ref{tab:Features}). These are standard features used
in information retrieval for web search, most of them described in
\citet{DBLP:journals/corr/QinL13}. Features related only to publication
are labelled as $P$ and features related to publication--grant pairs
are labelled as $\mathit{\mathit{\textrm{P-G}}}$. The annotations
are defined as below:

1. A publication description consists unique terms $q=\{q_{1},q_{2},...,q_{m}\}$.
We define the length of publication description$\left|q\right|$as
the number of tokens it contains, with $m\leq\left|q\right|$. Similarly,
we represent a grant description as $d=\{d_{1},d_{2},...,d_{n}\}$,
where the length $\left|d\right|$ is the number of tokens $d$ contains,
with $n\leq\left|d\right|$. We denote the corpus $D$ as the collection
of all the grant descriptions and $\left|D\right|$ as the total number
of grants in the corpus.

2. We use $c(q_{i},q)$ to denote the number of times a publication
token $q_{i}$ appears in a publication $q$. Similarly, we use $c(q_{i},d)$
to denote the number of times a publication token $q_{i}$ appears
in the grant $d$, and $c(q_{i},D)$ to denote the number of occurrences
of $q_{i}$ in the corpus $D$.

3. The terms frequency of a publication is denoted as $tf(q)$, the
document frequency $df(q_{i})$ is the number of grants containing
term $q_{i}$, and the inverse document frequency of a publication
term $q_{i}$ is denote as $idf(q_{i})$

4. The $LMIR$ features is a set of smoothing methods for estimating
the language model. The formal definition of these features is provided
in \citet{lmir}. For the Jelinek-Mercer smoothing method, we use
parameter $\lambda=0.1$. For smoothing using Dirichlet priors, we
set the parameter $\mu=2000$. For the Absolute Discount smoothing,
we use parameter $\text{\ensuremath{\delta=0.7}}$.

\begin{table}
\begin{centering}
\begin{tabular}{ll>{\raggedright}p{0.1\textwidth}}
\hline 
Feature \# & \textbf{Feature} & \textbf{Class}\tabularnewline
\hline 
1 & {\scriptsize$\sum_{q_{i}}c(q_{i},d)$ } & P-G\tabularnewline
2 & {\scriptsize$\sum_{q_{i}}log(c(q_{i},d)+1)$ } & P-G\tabularnewline
3 & {\scriptsize$\frac{\sum_{q_{i}}c(q_{i},d)}{\left|d\right|}$ } & P-G\tabularnewline
4 & $\left|d\right|$ & P\tabularnewline
5 & {\scriptsize$sum(tf(q))$} & P-G\tabularnewline
6 & {\scriptsize$min(tf(q))$} & P-G\tabularnewline
7 & {\scriptsize$max(tf(q))$} & P-G\tabularnewline
8 & {\scriptsize$mean(tf(q))$} & P-G\tabularnewline
9 & {\scriptsize$var(tf(q))$} & P-G\tabularnewline
10 & {\scriptsize$\frac{sum(tf(q))}{\left|g\right|}$} & P-G\tabularnewline
11 & {\scriptsize$\frac{min(tf(q))}{\left|g\right|}$} & P-G\tabularnewline
12 & {\scriptsize$\frac{max(tf(q))}{\left|g\right|}$} & P-G\tabularnewline
13 & {\scriptsize$\frac{mean(tf(q))}{\left|g\right|}$} & P-G\tabularnewline
14 & {\scriptsize$\frac{var(tf(q))}{\left|g\right|}$} & P-G\tabularnewline
15 & {\scriptsize$\sum_{q_{i}}log(\frac{\left|D\right|}{c(q_{i},D)+1}+1)$} & P\tabularnewline
16 & {\scriptsize$\sum_{q_{i}}idf(q_{i})$} & P\tabularnewline
17 & {\scriptsize$\sum_{q_{i}}log(idf(q_{i})+1)$} & P\tabularnewline
18 & {\scriptsize$sum(\text{c-idf}(q))$} & P-G\tabularnewline
19 & {\scriptsize$min(\text{c-idf}(q))$} & P-G\tabularnewline
20 & {\scriptsize$max(\text{c-idf}(q))$} & P-G\tabularnewline
21 & {\scriptsize$mean(\text{c-idf}(q))$} & P-G\tabularnewline
22 & {\scriptsize$var(\text{c-idf}(q))$} & P-G\tabularnewline
23 & {\scriptsize$sum(\text{weighted-c-idf}(q))$} & P-G\tabularnewline
24 & {\scriptsize$min(\text{weighted-c-idf}(q))$} & P-G\tabularnewline
25 & {\scriptsize$max(\text{weighted-c-idf}(q))$} & P-G\tabularnewline
26 & {\scriptsize$mean(\text{weighted-c-idf}(q))$} & P-G\tabularnewline
27 & {\scriptsize$var(\text{weighted-c-idf}(q))$} & P-G\tabularnewline
28 & {\scriptsize$BM25(q,d)$} & P-G\tabularnewline
29 & {\scriptsize$LMIR.AbsoluteDiscount$} & P-G\tabularnewline
30 & {\scriptsize$LMIR.Dirichlet$} & P-G\tabularnewline
31 & {\scriptsize$LMIR.Jelinek-Mercer$} & P-G\tabularnewline
\hline 
\multicolumn{3}{l}{\textit{Note: $tf(q)=\{c(q_{1},q),c(q_{2},q),\cdots,c(q_{m},q)\}$}}\tabularnewline
\multicolumn{3}{l}{\textit{$idf(q_{i})=\log(\frac{\left|D\right|}{df(q_{i})+1})$, where
$df(q_{i})$ is the number of grants containing term $q_{i}$ }}\tabularnewline
\multicolumn{3}{l}{\textit{$\text{c-idf}(q)=\{c(q_{k},g)\cdot idf(q_{k})\}_{k=1,\dots,m}$}}\tabularnewline
\multicolumn{3}{l}{\textit{$\text{weighted-c-idf}(q)=\{\frac{c(q_{k},d)}{\left|d\right|}\cdot idf(q_{k})\}_{k=1,\dots,m}$}}\tabularnewline
\multicolumn{3}{>{\raggedright}p{0.3\textwidth}}{\textit{$BM25(q,d)=\underset{q_{i}\in p}{\sum}\frac{idf(q_{i})\cdot c(q_{i},d)\cdot(k_{1}+1)}{c(q_{i},d)+k_{1}\cdot(1-b+b\cdot\frac{\left|d\right|}{avgdoclen})}$,
where $k_{1}=1.5$ and $b=0.75$, the $avgdoclen$ refer to the average
document length of the entire corpus. P: scientist's publication G:
grant.}}\tabularnewline
\end{tabular}
\par\end{centering}
\caption{\protect\label{tab:Features}Features of the system. }
\end{table}

\subsubsection{Semantic features}

In order to capture the semantic of the grant description and publication,
we make use of the distributed word representations. Inspired by the
idea ``you should know a word by the company it keeps'' proposed
by \citet{firth57synopsis}, there are a set of techniques committed
to represents word as a multi-dimensional vector of continuous real
numbers, each dimension captures a facet of the word's meaning, the
real number represent the strength of that meaning. Thus, the semantically
similar words are located close to each other in the vector geometric
space. The fastText \citep{bojanowski2016enriching} word vector is
one of the popular pre-trained word semantic representation. By using
the character level information, fastText achieves good performance
and is able to process the words which do not exist in the training
corpora. We obtained a copy of fastText vector trained on large scale
Common Crawl (web pages) and Wikipedia \citep{grave2018learning}.
Each vector has 300 dimensions.

We represent the description of a grant and a publication as vectors
by averaging the fastText vectors of each word they contain. Then
we use the cosine similarity between the grant and publication vectors
as semantic feature.

\section{Experiments and Result}

We first report the performance, then attempt to interpret what are
the features that the model considers important during matching.

\subsection{Evaluation Metric}

We use Normalized Discounted Cumulative Gain (NDCG) as our evaluation
metric. The NDCG is designed for non-binary relevance labels, and
usually evaluated over top k search results. The NDCG@k is defined
as, 

\begin{equation}
NDCG@k=\frac{\stackrel[i=1]{k}{\sum}\frac{(2^{\mathit{rel}_{i}}-1)}{\mathit{log_{2}(i+1)}}}{\stackrel[i=1]{k}{\sum}\frac{(2^{\mathit{ideal}_{i}}-1)}{\mathit{log_{2}(i+1)}}},
\end{equation}
where $k$ is a particular rank position, $\mathit{rel_{i}}$ is the
predicted relevance order at position $i$, the $\mathit{ideal_{i}}$
is the ideal relevance order (ground truth) at position $i$. The
value of NDCG ranges from 0 to 1, the higher the better.

\subsection{Ranking Function Performance}

As the user mainly pay attention to the top items on the recommended
list, we want the top items ranked by our model with less error. In
our experiment, we optimize the NDCG over top 1 and top 5 in the ranked
list. As it shown in Table \ref{tab:performance}, the NDCG@1 achieves
a score is 0.945 (the highest possible score is 1), the NDCG@5 are
very close to The NDCG@1, and both of them achieves a relatively high
score, indicating that GotFunding can rank the grant candidates well. 

\begin{table}
\begin{centering}
\begin{tabular}{cc}
\hline 
Metric & Performance\tabularnewline
\hline 
Training(NDCG@1) & 0.975\tabularnewline
Training(NDCG@5) & 0.954\tabularnewline
Validation(NDCG@1) & \textbf{0.945}\tabularnewline
Validation(NDCG@5) & 0.943\tabularnewline
\hline 
\end{tabular}
\par\end{centering}
\caption{\protect\label{tab:performance}The ranking function performance}
\end{table}

\subsection{Feature Importance}

One of the goal of our research is to understand what the most effective
factors matching publications to grants. Using \textsc{GotFunding},
we can begin to understand these factors by interpreting the importance
in the LambdaRank gradient boosting algorithm. lightGBM provides that
cumulative total grains of splits for features, which is commonly
used as feature importance \citep{lightgbm}.

\subsubsection{Feature importance analysis}

\begin{figure}
\begin{centering}
\includegraphics[width=1\textwidth]{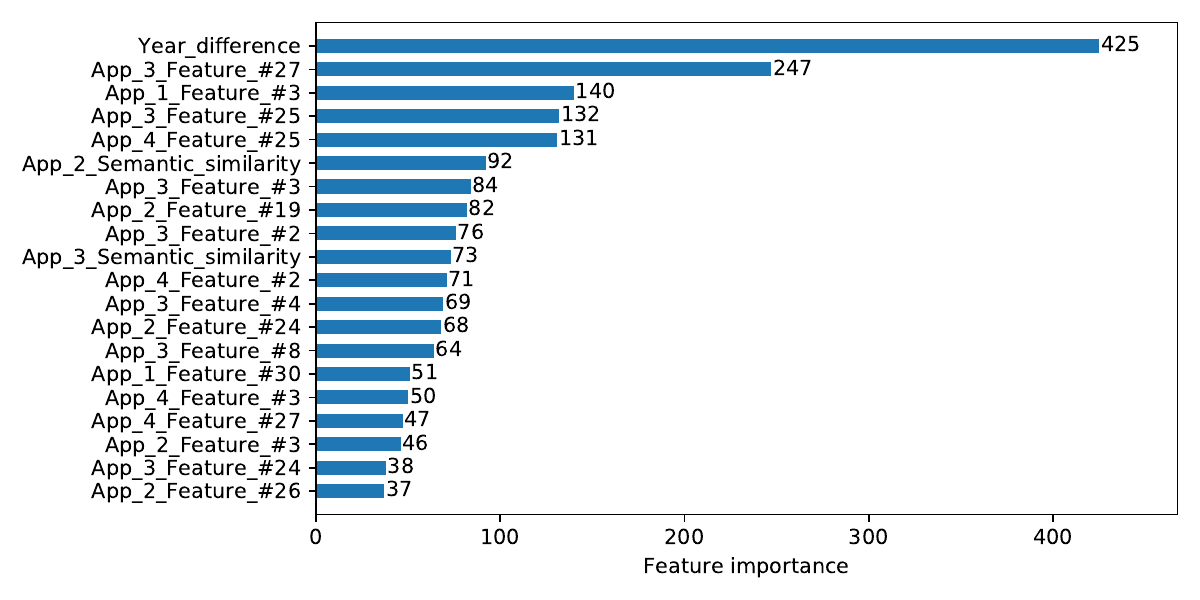}
\par\end{centering}
\caption{\protect\label{fig:Top-20-feature}Top 20 feature importance. The
APP\_{[}1,2,3,4{]} in the feature name denotes the four approaches
used for the grant description. The Feature\_\#{[}1-31{]} corresponds
to Table \ref{tab:Features}. The top three features are year difference
between publication and grant, information content of publication,
and relevance between publication and grant.}
\end{figure}

We listed the 20 most important features in Figure \ref{fig:Top-20-feature}.
The most import features is the temporal similarity between the publication
and the grant. Its importance is almost twice as the second most important
feature. The top second feature is the amount of information in the
publication represented by the size of the publication document. The
third feature and forth features are related to the relevance between
publication and grant. The top 10 features account for almost 50\%
of the overall importances. Taken together, these top three features
represent a combination of high relevance with high coverage.

Because our features are separated into four group of features, we
wanted to understand the importance of each of these subgroups. These
groups are funding agency information, the grant title, grant abstract
and their union. To do this, we computed the total feature importance
of all the features belonging to a group. Our analysis shows that
the importance rank the groups as follows: grant abstract (total importance:
1063), The combination of funding agency information, grant title
and grant abstract (total importance: 645), grant title (total importance:
440), and grant agency information(total importance: 427). These results
suggest that the abstract contributed the most.

We also analyzed the different importance of the semantic vs the statistical
features. We again do this by computing the total feature importance
of the features that belong to these two categories. Our results show
that the statistical features (total importance: 2791) is larger than
semantic features (total importance: 209). 

\section{Discussion and Conclusion}

In our work, we aim at improving how scientists can find relevant
grants based on their research profile. We propose to solve this problem
by building a recommendation system that learns from historical publication--grant
relationships. Our results show that we can achieve a high performance
of NDCG@1=0.975. Further, by analyzing what the recommendation system
learned we can estimate that the most successful links between a publication
and grant are when they are temporally relevant, the publication has
large amounts of information (e.g., long document), and there is a
good relevance between the publication and grant. We now discuss some
limitations and future work.

One of the limitation of our work is that we only look at grants that
were funded in the past. Funding mechanisms might be changing over
time and publication topics might also change over time. This means
that there is no guarantee that a correct prediction will actually
yield a successful match for a future grant. Recommendation systems
however benefits from large amounts of data and unless we are able
to interview and ask scientists about their opinion on publication
to grant matching, it is hard to build a recommendation system otherwise.
Finally, even if our recommendations are off by topic, they can still
serve as a narrowing step during the initial stages of match searchers. 

Another limitation of our work is that we are using publications that
we already funded by grants. However, our recommendation is trying
to solve the opposite problem whereas a scientists wants to find a
publication that can initiate funding. Research is still unclear on
whether funding changes the direction of research but even if it is
does, our recommendation could be useful to discover people that have
worked on similar problems to ours. We hope to obtain new data in
the future about grants that were not funded because they did not
meet the criteria of a certain problem. Thus, with data that is available
but potentially harder to obtain, some of these issues could be solved.

Our recommendation system is one of the first ones that offers scientists
the ability to match their research to past grants. We think this
research direction will benefit specially those who are starting in
their career and might not have the human capital to help in finding
relevant funding opportunities.

\bibliographystyle{apalike}
\bibliography{bibliography}

\end{document}